\documentstyle[pra,aps,eqsecnum,amssymb,amsmath,amsthm,twocolumn]{revtex}
\tighten
\begin{document}
\clearpage
\preprint{}
\draft

\title{Sine distance for quantum states}

\author{ Alexey E. Rastegin }
\address{Department of Theoretical Physics, Irkutsk State University,
Gagarin Bv. 20, Irkutsk 664003, Russia}

\maketitle
\begin{abstract}
We thoroughly analyse the distance between quantum states that has
been applied to state-dependent cloning and partly studied in the
previous work of the author [Phys. Rev. A {\bf 66}, 042304 (2002)].
Elementary proofs of its significant properties are given.
\end{abstract}


\pagenumbering{arabic}
\setcounter{page}{1}

\protect\section{Introduction}

Over the last twenty years there have been impressive theoretical and
experimental advances in actions on single quantum and use of them
for information processing. Above all, quantum cryptography
\cite{gisin}, quantum factoring \cite{shor} and quantum searching
\cite{grover} are most inspiring. Quantum devices can apparently
provide more powerful tools of communication and computation than
classical ones. A design of efficient algorithm for some quantum
information processing task demands that we compare results of tested
quantum operations. So mathematical techniques of quantum information
theory must include quantitative measures of closeness of quantum
states.

It would seem that for pure states the square-overlap provides such a
measure. But quantum information is a growing branch with many
facets. Natural as the square-overlap is, it is not able to give the
best measure in all respects. In Ref. \cite{rast1} the author offered
a new approach to the state-dependent cloning. As it is well-known
\cite{acin}, the majority of the studies of cloning uses the figures
of merit based on the fidelity. For pure states it is reduced to the
the square-overlap. The new figure of merit, called "relative error",
is based on {\it sine of angle between two states} \cite{rast1}. As
it turned out, this figure of merit is dualistic to those arising from
the square-overlap. In Ref. \cite{rast3} the above approach was extended
to mixed-state cloning. So the study of the relative error has allowed
us to complement the portrait of state-dependent cloning
\cite{rast1,rast3}.Thus, there may be more than one way to fit the
problem of state closeness, even if the states are pure.

We see that the problem of building a good distance for quantum states
is inevitable. Moreover, we rather need some set of reliable measures,
complementing each other. In fact, it is impossible to foresee all
potential questions, because many fields are still undeveloped. Here
several example may be pointed out. Since many results in the quantum
cloning used the fidelity, the writers of Ref. \cite{acin} have
stated such a question. What about other figures of merit of clones?
Reviewing the ideas of quantum information within the frames of
relativity theory, Peres and Terno rised a few important open
problems \cite{terno}. Of course, any workable notion of distance
must be bound in the natural way. First, a distance should have clear
physical meaning. Second, it should have a direct expression in the
terms of density matrices. At last, there is spesific "distance"
property: it must be a metric.

The aim of the present work is to further clarify understanding of
the sine of angle between two states as a distance measure. It will
be referred as to "sine distance". As it is shown in Refs.
\cite{rast1,rast3}, the use of the sine distance as distance measure
provides new and fruitful viewpoint on the state-dependent cloning.
More recently, the writers of Ref. \cite{nielsen} found that the
above distance naturally arises in the context of quantum
computation. But the sine distance is not previously studied as
independent notion. The contribution of the paper is to fill up this
lacuna. To simplify the exposition, we separately consider the case
of pure states. It is such a case that is natural. First, many key
ideas of quantum information were discovered in the models those deal only
with pure states. Second, the analysis of transformations of mixed
states demands more powerful techniques. At last, when the states are
impure, a qualitative restatement of considered task may be needed.
For example, for pure states the cloning is equivalent of
broadcasting, but for mixed states the cloning is very special case
of broadcasting \cite{barnum}. Starting with pure-state case, we then
examine the general case of mixed states. After briefly reviewing of
background material on impure states, we establish basic properties
of the sine distance. Further, the sine distance is considered in the
context of quantum operations. The principal role of this is to
provide the clear justification why the notion of sine distance is
reliable and convenient. In the present work by the quantum states we
mean that ones all are normalized.

\protect\section{The case of pure states}

In this section we shall give a clear definition of "sine
distance" for pure states. As every, we define the angle
$\delta(x,y)\in[0;\pi/2]$ between pure states $|x\rangle$ and
$|y\rangle$ by
\begin{equation}
\delta(x,y):=\arccos
\ \bigl| \langle x|y \rangle \bigr|
\ . \label{eq21}
\end{equation}
We will also write $\delta_{xy}$ in bulky expressions. States
$|x\rangle$ and $|y\rangle$ are indentical if and only if
$\delta_{xy}=0$.

{\bf Definition 1} {\it Sine distance between pure states $|x\rangle$
and $|y\rangle$ is defined by}
\begin{equation}
d(x,y):=\sin\delta(x,y) \ .
\label{eq22}
\end{equation}

The pure state is a ray in the Hilbert space. So, for given two states
$|x\rangle$ and $|y\rangle$ we can always suppose that $\langle
x|y\rangle$ is a nonnegative real number. Let us denote
\begin{align}
 & |x\rangle=\cos\theta|0\rangle+
\sin\theta|1\rangle \label{eq23} \ , \\
 & |y\rangle=\sin\theta|0\rangle+
\cos\theta|1\rangle \label{eq24} \ ,
\end{align}
that is customary in state-dependent cloning \cite{bruss}. The
vectors $|0\rangle$ and $|1\rangle$ are orthonormal and
$2\theta\in[0;\pi/2]$. So, the overlap between $|x\rangle$ and
$|y\rangle$ is $\,|\langle x|y \rangle|=\sin2\theta\,$, and by
Eqs. (\ref{eq21}) and (\ref{eq22}) we therefore have
\begin{equation}
d(x,y)=\cos2\theta \ .
\label{eq25}
\end{equation}
The convenience of parametrization by Eqs. (\ref{eq23}) and
(\ref{eq24}) is that for each linear operator $L$ we have
\begin{equation}
\langle x|L|x \rangle -
\langle y|L|y \rangle=
\bigl\{ \langle 0|L|0 \rangle
-\langle 1|L|1 \rangle\bigr\}\, d(x,y) \>.
\label{eq26}
\end{equation}

It turns out that for any quantum operation the quantity $d(x,y)$
estimates the difference between probabilities of processes begining
with inputs $|x\rangle$ and $|y\rangle$ respectively. Recall that a
quantum operation ${\cal{E}}$ is formal description of physical
process which starts with an input state $\sigma$ of quantum system
$S$ and results in an output state
$$
\sigma':=\frac{{\cal{E}}(\sigma)}{{\rm{tr}}\{{\cal{E}}(\sigma)\}}
$$
of (generally another) quantum system $S'$. The normalizing divisor
is trace over the Hilbert space ${\cal{H}}'$ of $S'$ and gives the
probability that such a process occurs. So we have
$0\leq{\rm{tr}}\{{\cal{E}}(\sigma)\}\leq1$. The domain of ${\cal{E}}$
is real vector space of Hermitian operators on the input space
${\cal{H}}$. The range of ${\cal{E}}$ is a subset of real vector
space of Hermitian operators on the output space ${\cal{H}}'$. It is
necessary that this map be linear and completely positive
\cite{kraus}.

The operator-sum representation is a key result of the theory of
quantum operations. That is \cite{kraus}, the map ${\cal{E}}$ is a
quantum operation if and only if
$$
{\cal{E}}(\sigma)=\sum\nolimits_{\mu} E_{\mu}\sigma
E_{\mu}^{\dagger}
$$
for some set of operators $\{E_{\mu}\}$. These operators map the input
space ${\cal{H}}$ to the output space ${\cal{H}}'$ and satisfy
\begin{equation}
\sum\nolimits_{\mu} E_{\mu}^{\dagger} E_{\mu} \leq {\mathbf{1}}
\ . \label{eq27}
\end{equation}
It is necessary for proper probabilistic treatment. The set
$\{E_{\mu}\}$ completely specifies a quantum operation. For given set
of such operators let ${\cal{E}}_{\nu}$ denotes a operation specified
by single operator $E_{\nu}\,$, i.e.
${\cal{E}}_{\nu}(\sigma):=E_{\nu}\sigma E_{\nu}^{\dagger}\,$.
According to the terminology of Ref. \cite{caves}, such a quantum
operation is contained in the class of {\it ideal} operations. The
following statement is a basic result of this section.

{\bf Proposition 1} {\it If the set $\{E_{\mu}\}$ of operators
specifies a quantum operation ${\cal{E}}$ then}
\begin{align}
& \left|\, {\rm{tr}}\bigl\{{\cal{E}}(|x\rangle\langle x|)\bigr\}
- {\rm{tr}}\bigl\{{\cal{E}}(|y\rangle\langle y|)\bigr\}
\right|\leq d(x,y) \label{eq28} \>, \\
\sum_{\mu} & \left|\,
{\rm{tr}}\bigl\{{\cal{E}}_{\mu}(|x\rangle\langle x|)\bigr\} -
{\rm{tr}}\bigl\{{\cal{E}}_{\mu}(|y\rangle\langle y|)\bigr\}
\right|\leq 2d(x,y) \label{eq29} \>.
\end{align}

{\bf Proof} Because the trace of sum of operators is equal to the sum
of traces of these operators, we obtain
\begin{equation}
{\rm{tr}}\bigl\{{\cal{E}}(|v\rangle\langle v|)\bigr\}
=\sum\nolimits_{\mu}
{\rm{tr}}\bigl\{{\cal{E}}_{\mu}(|v\rangle\langle v|)\bigr\}
\label{eq210}
\end{equation}
for $v=x,y$. Due to
$\,{\rm{tr}}\{|u'\rangle\langle v'|\}=\langle v'|u'\rangle\,$, each
term of the latter sum can be expressed as
\begin{equation}
{\rm{tr}}\bigl\{E_{\mu}|v\rangle\langle v|E_{\mu}^{\dagger}\bigr\}
=\langle v|{\mathbf{T}}_{\mu}|v\rangle
\ , \label{eq211}
\end{equation}
where $\,{\mathbf{T}}_{\mu}:=E_{\mu}^{\dagger} E_{\mu}\,$.
By Eqs. (\ref{eq210}) and (\ref{eq211}) we have
\begin{equation}
{\rm{tr}}\bigl\{{\cal{E}}(|v\rangle\langle v|)\bigr\}
=\langle v|{\mathbf{T}}|v \rangle
\ , \label{eq212}
\end{equation}
where positive operator
$\,{\mathbf{T}}:=\sum_{\mu}{\mathbf{T}}_{\mu}\,$. By Eqs.
(\ref{eq212}) and (\ref{eq26}), we rewrite the left-hand side of Eq.
(\ref{eq28}) as
\begin{equation}
\bigl| \langle 0|{\mathbf{T}}|0 \rangle
-\langle 1|{\mathbf{T}}|1 \rangle\bigr|\, d(x,y)\ ,
\label{eq213}
\end{equation}
By Eq. (\ref{eq27}), we get
${\mathbf{0}}\leq{\mathbf{T}}\leq{\mathbf{1}}$,
$0\leq\langle u|{\mathbf{T}}|u\rangle\leq1$ and
$$
-1\leq\langle 0|{\mathbf{T}}|0 \rangle
-\langle 1|{\mathbf{T}}|1 \rangle \leq +1\>.
$$
So due to expression (\ref{eq213}) we obtain Eq. (\ref{eq28}).

Continuing, due to Eqs. (\ref{eq211}) and (\ref{eq26}) we can write
\begin{align*}
 & \left|\,
{\rm{tr}}\bigl\{{\cal{E}}_{\mu}(|x\rangle\langle x|)\bigr\} -
{\rm{tr}}\bigl\{{\cal{E}}_{\mu}(|y\rangle\langle y|)\bigr\}
\right| \\
 & =\bigl|\, \langle 0|{\mathbf T}_{\mu}|0 \rangle -
\langle 1|{\mathbf T}_{\mu}|1 \rangle\, \bigr|\, d(x,y) \\
 & \leq \bigl\{ \langle 0|{\mathbf T}_{\mu}|0 \rangle
+\langle 1|{\mathbf T}_{\mu}|1 \rangle\bigr\}\, d(x,y) \ .
\end{align*}
To sum over all $\mu$'s, we see that the left-hand side of
Eq. (\ref{eq29}) is not larger than
$\{\langle 0|{\mathbf{T}}|0 \rangle+
\langle1|{\mathbf{T}}|1\rangle\}\,d(x,y)\,$. By
${\mathbf{0}}\leq{\mathbf{T}}\leq{\mathbf{1}}$, the latter does not
exceed $2d(x,y)$. $\blacksquare$

It must be stressed that the general upper bounds, given by
Proposition 1, are least and cannot be refinement. We shall now show
this fact. In the remainder of this section let ${\mathbf{P}}$ be an
operator such that ${\mathbf{0}}\leq{\mathbf{P}}\leq{\mathbf{1}}$ and
${\rm span}\{|x\rangle,|y\rangle\}$ is a subspace of its kernel.

{\bf Proposition 2} {\it If a quantum operation ${\cal{E}}$
reaches the first upper bound of Proposition 1 for given states
$|x\rangle$ and $|y\rangle$ then either
$\,{\mathbf{T}}=|0\rangle\langle0|+{\mathbf{P}}\,$ or
$\,{\mathbf{T}}=|1\rangle\langle1|+{\mathbf{P}}\,$.}

{\bf Proof} Take a basis $\{|j\rangle\}$ containing kets $|0\rangle$
and $|1\rangle$ from the parametrization by Eqs. (\ref{eq23}) and
(\ref{eq24}). As the proof of Proposition 1 shows, the left-hand side
of Eq. (\ref{eq28}) is equal to $|c_{00}-c_{11}|\,d(x,y)\,$, where
$\,c_{jk}:=\langle j|{\mathbf{T}}|k\rangle\,$. So the equality in Eq.
(\ref{eq28}) holds if and only if
\begin{equation}
|c_{00}-c_{11}|=1 \ .
\label{eq214}
\end{equation}
Due to ${\mathbf{0}}\leq{\mathbf{T}}\leq{\mathbf{1}}$ we have
$0\leq c_{jj}\leq1$ for all values of label $j$. Under the latter, Eq.
(\ref{eq214}) is satisfied in two cases: (i) $c_{00}=1$ and
$c_{11}=0$; (ii) $c_{00}=0$ and $c_{11}=1$. In the case (i) we obtain
\begin{equation}
{\mathbf{T}}=|0\rangle\langle 0|+
c_{01} |0\rangle\langle 1|+c_{10}|1\rangle\langle 0|+{\mathbf{P}}
\ , \label{eq215}
\end{equation}
where operator
\begin{equation}
{\mathbf{P}}:=\sum\nolimits_{j,k=2}^{N-1} c_{jk} |j\rangle\langle k|
\label{eq216}
\end{equation}
and $N:={\rm dim}({\cal{H}})$. The action of operator ${\mathbf{T}}$
in subspace ${\rm span}\{|x\rangle,|y\rangle\}$ is described by the matrix
\begin{equation}
\begin{pmatrix}
 1 & \alpha-i\beta \\
 \alpha+i\beta & 0
\end{pmatrix}
\ . \label{eq217}
\end{equation}
Here $\alpha$ and $\beta$ are real, and
$c^{*}_{01}=\alpha+i\beta=c_{10}$ by Hermisity. Due to positivity,
both eigenvalues of the matrix (\ref{eq217}) are nonnegative that
takes place if and only if $\alpha=\beta=0$. So $c_{01}=c_{10}=0$ and
from Eq. (\ref{eq215}) we obtain
$\,{\mathbf{T}}=|0\rangle\langle0|+{\mathbf{P}}\,$. By a parallel
argument, in the case (ii) we get
$\,{\mathbf{T}}=|1\rangle\langle1|+{\mathbf{P}}\,$. To satisfy
condition ${\mathbf{0}}\leq{\mathbf{T}}\leq{\mathbf{1}}$, the
operator ${\mathbf{P}}$ must obey
${\mathbf{0}}\leq{\mathbf{P}}\leq{\mathbf{1}}$. Due to definition by
Eq. (\ref{eq216}), ${\rm span}\{|x\rangle,|y\rangle\}$ is a subspace
of kernel of operator ${\mathbf{P}}$. $\blacksquare$

Just as the bound given by Eq. (\ref{eq28}), the upper bound given by
Eq. (\ref{eq29}) is also attainable. For example, the equality in Eq.
(\ref{eq29}) is reached by the quantum operation
$$
{\cal{E}}(|v\rangle\langle v|)=E_{0}|v\rangle\langle v|E_{0}^{\dagger}
+E_{1}|v\rangle\langle v|E_{1}^{\dagger} \ ,
$$
where two mapping operators satisfy
$E_{0}^{\dagger}E_{0}=|0\rangle\langle0|$ and
$E_{1}^{\dagger}E_{1}=|1\rangle\langle1|$. This follows from
Proposition 2. Since both bounds of Proposition 1 deal with
probabilities that corresponding process occurs, it provides
well-motived physical meaning of the sine distance.

However, the real devices are inevitably exposed to noise. Key result
of a quantum system interacting with its enviroment is the loss of
superposition, called "decoherence" \cite{zurek}. It is its action
that is to undo the interference of states used in data processing,
replacing them instead with mixtures of states. The solution of
subroutine problem required a consideration of quantum circuits with
mixed states \cite{aharonov}. As it is shown in Ref.  \cite{fan}, the
cloning machine, which can input any mixed state in symmetric
subspace, is necessary in quantum information. So, however hard the
careful examination of mixed states may be from the technique
viewpoint, we are to develop it. We shall now extend the notion of
sine distance to the case of mixed states.

\protect\section{The case of mixed states}

As it is well known, the square-overlap $|\langle x|y\rangle|^2$ is
the probability that $|y\rangle$ passes the yes/no test of "being the
state $|x\rangle$." This clear physical meaning gives us a better
understanding of why the square-overlap ensures a natural way to
distinguish pure states. However, there is no evident analog of
yes/no test for mixed states. Nevertheless, we can extend to mixed
states a few notions which are useful in the case of pure states.
This is provided by the concept of {\it purifications}.

According to the "decoherence" viewpoint \cite{zurek}, any mixed
state is describing the reduced states of a subsystem $S$ entangled
with the environment. The total system is being in a pure state. If
the quantum system $S$ is considered then we append system $Q$, which
is a copy of $S$. Widening the above viewpoint, we can imagine that a
mixed state $\sigma$ of $S$ arises by partial trace operation from
pure state of extended system $SQ$. Namely, there is a pure state
$|X\rangle$, called "purification", for which \cite{jozsa}
\begin{equation}
\sigma={\rm tr}_{Q}
\bigl\{|X\rangle\langle X|\bigr\}
\ . \label{eq31}
\end{equation}
For any mixed states its purification can be made, and for given one
such a pure state is not unique \cite{jozsa}.

In Ref. \cite{rast2} we have defined the angle
$\Delta(\sigma,\rho)\in[0;\pi/2]$ between mixed states $\sigma$ and
$\rho$ by
\begin{equation}
\Delta(\sigma,\rho)
:=\min \delta(X,Y) \ ,
\label{eq32}
\end{equation}
where the minimum is taken over all purifications $|X\rangle$ of
$\sigma$ and $|Y\rangle$ of $\rho$. The properties of angle between
mixed states are listed in Ref. \cite{rast2}. In particular, we have
\begin{equation}
\Delta(\sigma,\rho)\leq
\Delta(\sigma,\omega)+\Delta(\rho,\omega)
\ . \label{eq33}
\end{equation}
We are now able to extend the notion of "sine distance" to the case
of mixed states.

{\bf Definition 2} {\it Sine distance between mixed states $\sigma$
and $\rho$ is defined by}
\begin{equation}
d(\sigma,\rho):=\sin\Delta(\sigma,\rho) \ .
\label{eq34}
\end{equation}

The sine distance can simply be expressed in terms of fidelity
function. Recall that the fidelity function generalizes the
square-overlap. More precisely, for given mixed states $\sigma$ and
$\rho$ of system $S$ the fidelity is defined as
\begin{equation}
F(\sigma,\rho):=
\max\ \bigl|\langle X|Y \rangle\bigr|^2
\ , \label{eq35}
\end{equation}
where the maximum is taken over all purifications $|X\rangle$
of $\sigma$ and $|Y\rangle$ of $\rho$ \cite{jozsa}. Using Eqs.
(\ref{eq32}), (\ref{eq34}) and (\ref{eq35}), it is easy to verify that
\begin{equation}
d(\sigma,\rho)=
\sqrt{1-F(\sigma,\rho)}
\ . \label{eq36}
\end{equation}
The definition by Eq. (\ref{eq35}) gives a kind of physical meaning of
the fidelity and makes many of its properties to be clear. But this
formula does not provide a calculational tool for evaluating the
fidelity function. Fortunately, the notion of fidelity is equivalent
to Uhlmann's "transition probability" \cite{jozsa}. By Uhlmann's
"transition probability" formula \cite{uhlmann}, the fidelity of
states $\sigma$ and $\rho$ is
\begin{equation}
F(\sigma,\rho)=\left\{{\rm tr}_S
\Bigl[(\sqrt{\sigma}\rho\sqrt{\sigma})^{1/2}\Bigr]
\right\}^2 \ .
\label{eq37}
\end{equation}
So, the fidelity and, therefore, the sine distance can directly be
expressed in the terms of density operators.

The basic properties of the sine distance can be obtained from the
definition by Eq. (\ref{eq34}) and features of the fidelity function.
In particular, the sine distance is a metric. These properties are
stated by the following.

{\bf Theorem 1} {\it Sine distance ranges between 0 and 1, and
$d(\sigma,\rho)=0$ if and only if $\sigma=\rho$. It is symmetric,
i.e. $d(\sigma,\rho)=d(\rho,\sigma)$. It obeies the triangle
inequality:}
\begin{equation}
d(\sigma,\rho)\leq d(\sigma,\omega)+d(\rho,\omega)
\ . \label{eq38}
\end{equation}
{\it Its square is convex: if $q,r\geq0$ and $q+r=1$ then}
\begin{equation}
d^2(\sigma,q\rho+r\omega)\leq
q\,d^2(\sigma,\rho)+r\,d^2(\sigma,\omega)
\ . \label{eq39}
\end{equation}

{\bf Proof} The first and second properties are corollaries of
Definition 2. So we will prove only the triangle inequality and that
the square of sine distance is convex. To establish Eq. (\ref{eq38}),
we consider two cases:
\begin{align}
 & 0\leq \Delta(\sigma,\omega)+\Delta(\rho,\omega)\leq\pi/2
\ ; \label{eq310} \\
 & \pi/2\leq \Delta(\sigma,\omega)+\Delta(\rho,\omega)\leq\pi
\ . \label{eq311}
\end{align}
By definition, the angle lies in the range $[0;\pi/2]$, where the sine
is a nondecreasing function. Due to Eq. (\ref{eq33}), in the case of
Eq. (\ref{eq310}) we obtain
\begin{align*}
\sin\Delta_{\sigma\rho} & \leq
\sin\Delta_{\sigma\omega}\cos\Delta_{\rho\omega}+
\cos\Delta_{\sigma\omega}\sin\Delta_{\rho\omega} \\
& \leq
\sin\Delta_{\sigma\omega}+\sin\Delta_{\rho\omega} \ .
\end{align*}
Thus, in the case Eq. (\ref{eq310}) the triangle inequality for sine
distance is corollary of the one for angle. But in the case Eq.
(\ref{eq311}) it is not so! Here an independent proof is wanted. This
need is met by Lemma of Appendix A. To prove Eq. (\ref{eq39}), we
shall use the concavity of fidelity. That is \cite{jozsa}, for
$q,r\geq0$ and $q+r=1$ there holds
\begin{equation}
F(\sigma,q\rho+r\omega)\geq qF(\sigma,\rho)+rF(\sigma,\omega)
\ . \label{eq312}
\end{equation}
Due to Eq. (\ref{eq36}) the latter can be rewritten as
$$
1-d^2(\sigma,q\rho+r\sigma)\geq q\left\{1 -
d^2(\sigma,\rho)\right\}+r\left\{1-d^2(\sigma,\omega)\right\} .
$$
By $q+r=1$, the latter provides Eq. (\ref{eq39}). $\blacksquare$

It is not incurious that in some cases the sine distance shows
concavity. Namely, for each $\sigma=|x\rangle\langle x|$ we have
\begin{equation}
d(\sigma,q\rho+r\omega)\geq
qd(\sigma,\rho)+rd(\sigma,\omega)
\> . \label{eq313}
\end{equation}
Indeed, the fidelity function of states $|x\rangle\langle x|$ and
$\rho$ is equal to $\langle x|\rho|x\rangle$ \cite{jozsa}, whence
\begin{equation}
F(\sigma,q\rho+r\omega)=
qF(\sigma,\rho)+rF(\sigma,\omega)
\ . \label{eq314}
\end{equation}
Due to the Jensen's inequality for concave function,
$$
\sqrt{1-(q\zeta+r\xi)}\geq q\sqrt{1-\zeta}+r\sqrt{1-\xi} \ .
$$
Substituting $\zeta=F(\sigma,\rho)$ and $\xi=F(\sigma,\omega)$ to the
latter inequality, by Eqs. (\ref{eq36}) and (\ref{eq314}) we obtain
Eq. (\ref{eq313}). Note that for correctness of the above argument
the equality in Eq. (\ref{eq312}) is necessary.

Thus, the sine distance has useful properties. It ranges between 0
and 1, it is a metric on quantum states, and its square is convex. We
shall now consider the sine distance within the frames of quantum
operations.

\section{On the quantum operations}

We shall now extend the main result of Sect. II to the case of mixed
states. The concept of purification  provides a direct way to do
this.

{\bf Theorem 2} {\it If the set $\{E_{\mu}\}$ of operators specifies a
quantum operation ${\cal{E}}$ then}
\begin{align}
& \left|\, {\rm{tr}}\bigl\{{\cal{E}}(\sigma)\bigr\}
- {\rm{tr}}\bigl\{{\cal{E}}(\rho)\bigr\}
\right|\leq d(\sigma,\rho) \label{eq41} \>, \\
\sum\nolimits_{\mu} & \left|\,
{\rm{tr}}\bigl\{{\cal{E}}_{\mu}(\sigma)\bigr\} -
{\rm{tr}}\bigl\{{\cal{E}}_{\mu}(\rho)\bigr\}
\right|\leq 2d(\sigma,\rho) \label{eq42} \>.
\end{align}

{\bf Proof} Let us define new operators
\begin{equation}
G_{\mu}:=E_{\mu}\otimes{\mathbf{1}}_Q
\ , \label{eq43}
\end{equation}
those map the space ${\cal{H}}\otimes{\cal{H}}$ to the space
${\cal{H}}'\otimes{\cal{H}}$. Due to Eq. (\ref{eq27}), these
operators satisfy
\begin{equation}
\sum\nolimits_{\mu} G_{\mu}^{\dagger} G_{\mu}
\leq {\mathbf{1}}_{SQ}
\ . \label{eq44}
\end{equation}
So the set $\{G_{\mu}\}$ specifies a quantum operation
${\cal{G}}$ with input space ${\cal{H}}\otimes{\cal{H}}$ and output
space ${\cal{H}}'\otimes{\cal{H}}$. Take purifications $|X\rangle$ of
$\sigma$ and $|Y\rangle$ of $\rho$ such that
$d(\sigma,\rho)=d(X,Y)$.
As it is shown in Appendix B, we then have
\begin{align}
{\rm{tr}}_{S{\pmb{'}}}\bigl\{{\cal{E}}(\sigma)\bigr\} & =
{\rm{tr}}_{S{\pmb{'}}Q}\bigl\{{\cal{G}}(|X\rangle
\langle X|)\bigr\} \ , \label{eq45} \\
{\rm{tr}}_{S{\pmb{'}}}\bigl\{{\cal{E}}(\rho)\bigr\} & =
{\rm{tr}}_{S{\pmb{'}}Q}\bigl\{{\cal{G}}(|Y\rangle
\langle Y|)\bigr\} \ . \nonumber
\end{align}
Applying Eq. (\ref{eq28}) to operation ${\cal{G}}$, by the last two
equalities and $d(\sigma,\rho)=d(X,Y)$ we obtain Eq. (\ref{eq41}).

According to Appendix B, we also have
\begin{align}
{\rm{tr}}_{S{\pmb{'}}}\bigl\{{\cal{E}}_{\mu}(\sigma)\bigr\} & =
{\rm{tr}}_{S{\pmb{'}}Q}\bigl\{{\cal{G}}_{\mu}(|X\rangle
\langle X|)\bigr\} \ , \label{eq46} \\
{\rm{tr}}_{S{\pmb{'}}}\bigl\{{\cal{E}}_{\mu}(\rho)\bigr\} & =
{\rm{tr}}_{S{\pmb{'}}Q}\bigl\{{\cal{G}}_{\mu}(|Y\rangle
\langle Y|)\bigr\} \ , \nonumber
\end{align}
where ideal operation ${\cal{G}}_{\nu}$ is specified by single
operator $G_{\nu}\,$. Applying Eq. (\ref{eq29}) to all quantum
operations ${\cal{G}}_{\mu}$'s, by a parallel argument we obtain Eq.
(\ref{eq42}). $\blacksquare$

The measurement is an important type of quantum operation. In this
case the input and output spaces are the same. As pointed out by
Everett \cite{everett}, a general treatment of all observations by
the method of projection operators is impossible. The most general
quantum measurement is called a {\it positive operator valued
measure}, or POVM \cite{peres}. A POVM with $M$ distinct outcomes is
specified by a set of $M$ positive operators ${\mathbf{A}}_{\mu}$
obeying
\begin{equation}
\sum\nolimits_{\mu=1}^{M} {\mathbf A}_{\mu}={\mathbf{1}} \ .
\label{eq47}
\end{equation}
Note that the number $M$ of different outcomes is not limited above
by the dimensionality $N$, in contrast to von Neumann measurement. If
the system $S$ is prepared in state $\sigma$, then the probability of
$\mu$'th outcome is \cite{peres}
\begin{equation}
p_{\mu}(\sigma)
:={\rm tr}\left\{\sigma{\mathbf A}_{\mu}\right\}  \ .
\label{eq48}
\end{equation}
With each POVM element ${\mathbf{A}}_{\nu}$ one can associate an
ideal quantum operation ${\cal{A}}_{\nu}$ defined by
$$
{\cal{A}}_{\nu}(\sigma):=\sqrt{{\mathbf{A}}_{\nu}}
\,\sigma\,\sqrt{{\mathbf{A}}_{\nu}} \ .
$$
Due to the cyclic property of the trace, we then have
\begin{equation}
p_{\nu}(\sigma):=
{\rm tr}\bigl\{{\cal{A}}_{\nu}(\sigma)\bigr\} \ .
\label{eq49}
\end{equation}
Let us define also an operation
${\cal{A}}(\sigma):=\sum_{\mu}{\cal{A}}_{\mu}(\sigma)$. By Eq.
(\ref{eq47}), this quantum operation is trace-preserving, that is
$\,{\rm tr}\{{\cal{A}}(\sigma)\}=1\,$. Applying Eq. (\ref{eq41}) to
separately taken operation ${\cal{A}}_{\mu}$ and Eq. (\ref{eq42})
to trace-preserving operation ${\cal{A}}$, we obtain the following
result.

{\bf Corollary} {\it For arbitrary POVM there holds}
\begin{align}
& \bigl|\,p_{\mu}(\sigma)-p_{\mu}(\rho)
\bigr|\leq d(\sigma,\rho) \label{eq410} \>, \\
\sum\nolimits_{\mu=1}^M & \bigl|\,p_{\mu}(\sigma)
-p_{\mu}(\rho)\bigr|\leq 2d(\sigma,\rho) \label{eq411} \>.
\end{align}

Thus, if the sine distance $d(\sigma,\rho)$ is small then probability
distributions generated by states $\sigma$ and $\rho$ for any
measurement are close to each other. Note that special cases of Eq.
(\ref{eq410}) were proven in Refs. \cite{rast1,rast3}.

The trace-preserving operation also is an important type of quantum
operation. Considering the quantum circuits with mixed states, the
writers of Ref. \cite{aharonov} showed that a general quantum gate
performs trace-preserving, completely positive linear map. So it is a
trace-preserving operation. Recall that a quantum operation is
trace-preserving when the equality in Eq. (\ref{eq27}) holds, and so
for any state $\sigma$ we have $\,{\rm tr}\{{\cal{E}}(\sigma)\}=1\,$.

As it is known \cite{barnum}, the fidelity function cannot decrease
under any trace-preserving quantum operation. Due to Eq.
(\ref{eq36}), the sine distance cannot increase under any
trace-preserving operation. That is, if quantum operation ${\cal{E}}$
is trace-preserving then
\begin{equation}
d\bigl({\cal{E}}(\sigma),{\cal{E}}(\rho)\bigr)
\leq d(\sigma,\rho) \ . \label{eq412}
\end{equation}
When operation is not trace-preserving, the contrary inequality can
be valid. The quantum state separation is an evident example of such
an operation. In the special case of two inputs, the success outcome
of separation leads to decrease of the fidelity of two possible state
of the system \cite{feng}. So the sine distance will be increased.

As it is shown in Ref. \cite{rast3}, such an inequality holds:
$$
\left| F(\sigma,\omega) -
F(\rho,\omega) \right| \leq d(\sigma,\rho) \ .
$$
This, when combined with Eq. (\ref{eq412}), gives the following. If
the operation ${\cal{E}}$ is trace-preserving then for arbitrary
$\sigma,\rho\in{\cal{H}}_S$ and $\omega'\in{\cal{H}}'_S$ we have
\begin{equation}
\Bigl| F\bigl({\cal{E}}(\sigma),\omega'\bigr) -
F\bigl({\cal{E}}(\rho),\omega'\bigr) \Bigr| \leq
d(\sigma,\rho) \ .
\label{eq413}
\end{equation}
Thus, if the sine distance $d(\sigma,\rho)$ between inputs $\sigma$
and $\rho$ is small then the fidelities
$F({\cal{E}}(\sigma),\omega')$ and $F({\cal{E}}(\rho),\omega')$ are
nearly equal to each other. So for any choice of standard $\omega'$
the outputs ${\cal{E}}(\sigma)$ and ${\cal{E}}(\rho)$ will be poorly
distinguishable. In fact, a natural measure of distinction for mixed
states is provided by the fidelity function \cite{jozsa}. It is for
this reason that the above interpretation of Eq. (\ref{eq413}) is to
be preferred.

To sum up, we can say that the sine distance between two quantum
states provides a reliable measure of their closeness. As the results
of this section show, if the value of $d(\sigma,\rho)$ is small then
observable effects caused by states $\sigma$ and $\rho$ will be close
to each other. It should be pointed out that the relations derived
here can be useful in various contexts.

\protect\section{Conclusion}

We have examined the sine distance for general quantum states and
showed the reasons for its use. This distance measure has good formal
properties. Namely, it is a metric on quantum states and ranges
between 0 and 1, its square is convex. If the sine distance between
two states is known then we can estimate the difference between
experimental manifestations of these states. Moreover, this measure
cannot increase under any trace-preserving quantum operation. So in a
single step of quantum computation the distance between outputs does
not exceed the distance between inputs.

In addition to the angle and the sine distance, the Bures metric is 
also used \cite{hubner}. As it is well known, this metric is equal 
to the square root of the quantity $(2-2\sqrt{F})$, where $F$ denotes 
the fidelity. Note that the mentioned metrics all are closely related 
to each other. Which of these three distances is most preferable? One 
may scarcely maintain that such a formulation of question is 
justified. Rather, some distance should be preferred in a first kind 
of tasks, other distance should be preferred in a second kind of
tasks, and so on. For example \cite{rast4}, the use of angles in
calculations clarifies the origins of bounds for state-dependent
cloning, even if a merit of cloning is measured by the global
fidelity.

Nevertheless, the following must be emphasized. The sine distance
lies in the interval $[0;1]$, whereas the angle lies in $[0;\pi/2]$
and the Bures measure lies in $[0;\sqrt{2}]$. As the range of
distance values, the interval $[0;1]$ seems more natural. In
addition, the sine distance between two states allows to estimate
directly a distinction between their observable effects. So the sine
distance is a reliable measure of closeness of quantum states.

\appendix

\section{lemma}

{\bf Lemma} {\it If $\alpha,\beta\in[0;\pi/2]$ and
$\pi/2\leq\alpha+\beta\leq\pi$ then}
\begin{equation}
\sin\alpha+\sin\beta\geq 1 \ .
\label{aa1}
\end{equation}

{\bf Proof} At first, it should be pointed out that
\begin{equation}
\sin\alpha+\sin\beta\geq
\sin^2\alpha+\sin^2\beta
\label{aa2}
\end{equation}
due to $\alpha,\beta\in[0;\pi/2]$. Applying the usual trigonometry,
the right-hand side of Eq. (\ref{aa2}) can be rewritten as
$$
\frac{1-\cos2\alpha}{2}+\frac{1-\cos2\beta}{2}=
1-\cos(\alpha+\beta)\cos(\alpha-\beta) \>.
$$
By conditions $\alpha,\beta\in[0;\pi/2]$ and
$\pi/2\leq\alpha+\beta\leq\pi$, we have
$\,\cos(\alpha+\beta)\cos(\alpha-\beta)\leq0\,$ and so Eq.
(\ref{aa1}). $\blacksquare$

\section{Rewriting traces}

Let the state $\sigma$ has the spectral decomposition
$$
\sigma=\sum\nolimits_j \lambda_j
\> |a_j\rangle\langle a_j| \ .
$$
Due to the properties of tracing and Eq. (\ref{eq211}),
\begin{align}
{\rm{tr}}_{S{\pmb{'}}}\bigl\{{\cal{E}}_{\mu}(\sigma)\bigr\}
 & = \sum\nolimits_j \lambda_j \>
{\rm{tr}}_{S{\pmb{'}}}\bigl\{{\cal{E}}_{\mu}
(|a_j\rangle\langle a_j|)\bigr\} \nonumber\\
 & = \sum\nolimits_j \lambda_j \>
\langle a_j|{\mathbf{T}}_{\mu}|a_j\rangle
\ . \label{bb1}
\end{align}
Applying Eq. (\ref{eq210}) to the latter relation and summing over
all $\mu$'s, we have
\begin{equation}
{\rm{tr}}_{S{\pmb{'}}}\bigl\{{\cal{E}}(\sigma)\bigr\}
=\sum\nolimits_j \lambda_j \>
\langle a_j|{\mathbf{T}}|a_j\rangle
\ . \label{bb2}
\end{equation}

In terms of Schmidt polar form \cite{jozsa1}, any purification
$|X\rangle$ of $\sigma$ can be written as
$$
|X\rangle=\sum\nolimits_j \sqrt{\lambda_j}\
|a_j\rangle\otimes|f_j\rangle \ ,
$$
where kets $|f_j\rangle$ form an orthonormal set in ${\cal{H}}$.
Drawing clear analogy with Eqs. (\ref{eq211}) and (\ref{eq212}), we
can write
\begin{align}
{\rm{tr}}_{S{\pmb{'}}Q}\bigl\{{\cal{G}}_{\mu}
(|X\rangle\langle X|)\bigr\} & =\langle X|{\mathbf{L}}_{\mu}|X\rangle
\ , \label{bb3} \\
{\rm{tr}}_{S{\pmb{'}}Q}\bigl\{{\cal{G}}
(|X\rangle\langle X|)\bigr\} & =\langle X|{\mathbf{L}}|X\rangle
\ , \label{bb4}
\end{align}
where ${\mathbf{L}}_{\mu}:=G_{\mu}^{\dagger}G_{\mu}$ and
${\mathbf{L}}:=\sum_{\mu} {\mathbf{L}}_{\mu}$ act in the Hilbert space
${\cal{H}}\otimes{\cal{H}}$. According to the definition by Eq.
(\ref{eq43}),
$$
G_{\mu}|X\rangle=\sum\nolimits_j \sqrt{\lambda_j}
\ E_{\mu}|a_j\rangle\otimes|f_j\rangle \ .
$$
The overlap of this vector with the self is equal to
\begin{align}
\langle X|{\mathbf{L}}_{\mu}|X\rangle
 & = \sum\nolimits_{jk} \sqrt{\lambda_j\lambda_k} \
\langle a_j|E_{\mu}^{\dagger}E_{\mu}|a_k\rangle\langle f_j|f_k\rangle
\nonumber\\
 & = \sum\nolimits_j \lambda_j \>
\langle a_j|{\mathbf{T}}_{\mu}|a_j\rangle
\> , \label{bb5}
\end{align}
because the set $\{|f_j\rangle\}$ is orthonormal. Due to Eq.
(\ref{bb5}), the left-hand side of Eq. (\ref{bb1}) is equal to the
left-hand side of Eq. (\ref{bb3}), and so we obtain Eq.
(\ref{eq46}).

Summing Eq. (\ref{bb5}) over all $\mu$'s, we at once obtain
$$
\langle X|{\mathbf{L}}|X\rangle=
\sum\nolimits_j \lambda_j \>
\langle a_j|{\mathbf{T}}|a_j\rangle \ .
$$
Therefore, the left-hand side of Eq. (\ref{bb2}) is equal to the
left-hand side of Eq. (\ref{bb4}), and so we get Eq. (\ref{eq45}).


\begin{references}

\bibitem{gisin}
N.~Gisin, G.~Ribordy, W.~Tittel and H.~Zbinden, Rev. Mod. Phys.
{\bf 74}, 145 (2002)

\bibitem{shor}
P.~W.~Shor, In {\it Proceedings of the 35th Annual Symposium on
Foundations of Computer Sciense} (IEEE Computer Society Press, New
York, 1994), pp. 124--134; D.~Beckman, A.~N.~Chari, Sr.~Devabhaktuni
and J.~Preskill, Phys. Rev. A {\bf 54}, 1034 (1996)

\bibitem{grover}
L.~K.~Grover, Phys. Rev. Lett. {\bf 79}, 325 (1997); M.~Boyer,
G.~Brassard, P.~Hoyer and A.~Tapp, Fortsch. Phys. {\bf 46}, 493
(1998); E.~Biham, O.~Biham, D.~Biron, M.~Grassl, D.~A.~Lidar and
D.~Shapira, Phys. Rev. A {\bf 63}, 012310 (2001)

\bibitem{rast1}
A.~E.~Rastegin, Phys. Rev. A {\bf 66}, 042304 (2002)

\bibitem{acin}
V.~Scarani, S.~Iblisdir, N.~Gisin and A.~Acin, Rev. Mod. Phys.
{\bf 77}, 1225 (2005)

\bibitem{rast3}
A.~E.~Rastegin, J. Opt. B: Quantum Semiclassical Opt. {\bf 5}, 647
(2003); A.~E.~Rastegin, arXiv: quant-ph/0208159

\bibitem{terno}
A.~Peres and D.~R.~Terno, Rev. Mod. Phys. {\bf 76}, 93 (2004)

\bibitem{nielsen}
A.~Gilchrist, N.~K.~Langford and M.~A.~Nielsen, Phys. Rev. A {\bf
71}, 062310 (2005)

\bibitem{barnum}
H.~Barnum, C.~M.~Caves, C.~A.~Fuchs, R.~Jozsa and B.~Schumacher,
Phys. Rev. Lett. {\bf 76}, 2818 (1996)

\bibitem{bruss}
D.~Bru{\ss}, D.~P.~DiVincenzo, A.~Ekert, C.~A.~Fuchs, C.~Macchiavello
and J.~A.~Smolin, Phys. Rev. A {\bf 57}, 2368 (1998)

\bibitem{kraus}
K.~Kraus, {\it States, Effects and Operations: Fundamental Notions of
Quantum Theory} (Springer-Verlag, Berlin, 1983)

\bibitem{caves}
M.~A.~Nielsen and C.~M.~Caves, Phys. Rev. A {\bf 55}, 2547 (1997)

\bibitem{zurek}
W.~H.~Zurek, Rev. Mod. Phys. {\bf 75}, 715 (2003)

\bibitem{aharonov}
D.~Aharonov, A.~Kitaev and N.~Nisan, arXiv: quant-ph/9806029

\bibitem{fan}
H.~Fan, Phys. Rev. A {\bf 68}, 052301 (2003)

\bibitem{jozsa}
R.~Jozsa, J. Mod. Optics {\bf 41}, 2315 (1994)

\bibitem{rast2}
A.~E.~Rastegin, Phys. Rev. A {\bf 67}, 012305 (2003)

\bibitem{uhlmann}
A.~Uhlmann, Rep. Math. Phys. {\bf 9}, 273 (1976)

\bibitem{everett}
H. Everett, Rev. Mod. Phys. {\bf 29}, 454 (1957)

\bibitem{peres}
A.~Peres, {\it Quantum Theory: Concepts and Methods} (Kluwer,
Dordrecht, 1993)

\bibitem{feng}
A.~Chefles and S.~M.~Barnett, J. Phys. A: Math. Gen. {\bf 31}, 10097
(1998); Y.~Feng, R.~Duan and Z.~Ji, Phys. Rev. A {\bf 72}, 012313
(2005)

\bibitem{hubner}
M.~H\"{u}bner Phys. Lett. A {\bf 163}, 239 (1992); M.~H\"{u}bner
Phys. Lett. A {\bf 179}, 226 (1993)

\bibitem{rast4}
A.~E.~Rastegin, Phys. Rev. A {\bf 68}, 032303 (2003); A.~E.~Rastegin,
arXiv: quant-ph/0301132

\bibitem{jozsa1}
L.~P.~Hughston, R.~Jozsa and W.~K.~Wootters, Phys. Lett. A {\bf 183},
14 (1993)

\end{references}
\end{document}